\documentclass[letterpaper]{article}
\usepackage{indentfirst}

\usepackage[T1]{fontenc}

\usepackage{geometry}
\usepackage{setspace}

\usepackage{achemso}

\usepackage{graphicx}
\usepackage{float}
\usepackage{multirow}
\usepackage{rotating}
\newfloat{scheme}{htbp}{los}
\floatname{scheme}{Scheme}
\floatname{chart}{Chart}
\newfloat{graph}{htbp}{loh}

\usepackage{chemformula} % Formulas using \ch{}
\usepackage[version = 4]{mhchem} % Formulas using \ce{}

\usepackage{authblk}
\author[1]{Kosuke Takiguchi*}
\author[1]{Yoshiharu Krockenberger}
\author[2]{Eisuke Magome}
\author[1]{Yoji Kunihashi}
\author[1]{Hideki Yamamoto}

\affil[1]{Basic Research Laboratories, NTT Inc., 3-1 Morinosato-Wakamiya, Atsugi, Kanagawa 243-0198, Japan}

\affil[2]{SAGA Light Source, 8-7 Yayoigaoka, Tosu, Saga 841-0005, Japan}

\title{Molecular beam epitaxy synthesis of ternary nitride \ch{PrTaN2} and its crystal structure determination}
\date{*Email: kosuke.takiguchi@ntt.com}

\begin{document}

\maketitle

\begin{abstract}
We report the discovery of a novel ternary nitride \ch{PrTaN2} synthesized as a thin film using molecular beam epitaxy. The combination of e-beam evaporation for refractory elements and a radio-frequency nitrogen radical source enables growth under a highly nitriding environment, providing access to phases not readily obtained in bulk synthesis. Structural characterization by X-ray diffraction and high-angle annular dark-field scanning transmission electron microscopy reveals that the compound crystallizes in an orthorhombic structure and grows with a well-defined orientation on \ch{YAlO3} substrates, while remaining essentially strain-free. To determine the crystal structure from limited thin-film diffraction data, we developed a fitting procedure based on structure factors. By combining extinction rules with constraints from Wyckoff positions, the number of fitting parameters is significantly reduced, enabling reliable structure determination. Systematic exclusion of alternative candidate phases in the Pr-Ta-N system, together with structure factor fitting, identifies the space group as $P222$ and determines the atomic coordinates. The present results demonstrate that thin-film growth with molecular beam epitaxy can stabilize previously unexplored ternary nitrides, and establish a practical approach for structural determination in such systems. This work provides a pathway for the exploration of new complex nitride materials.
\end{abstract}

\section*{Keywords}

Molecular beam epitaxy, X-ray diffraction, ternary nitride.

\section{Introduction}
Ternary complex nitrides with lanthanides are emerging and attracting attention as materials with rich physical properties \cite{Sun2019,Niewa1998,DiSalvo1996}. The complex nitrides composed of lanthanides and group V--VII 5$d$ transition metals (Ta, W, and Re) exhibit perovskite and Ruddlesden--Popper crystal structures \cite{Cario2001,Choi2025,Talley2021,Klo2021,Weidemann2024,Zakutayev2024}. Analogous to oxide perovskite and Ruddlesden--Popper phases, these ternary nitrides are expected to show superconductivity, ferromagnetism, and ferroelectricity. However, despite many theoretical predictions of stable ternary nitrides \cite{Grosso2023,Flores-Livas2019,Sarmiento-Prez2015}, the number of experimentally realized compounds remains limited. One major challenge is controlling group V--VII 5$d$ transition metals during synthesis. The melting points of these elements exceed 3000\,$^\circ$C, making them difficult to handle. Another challenge is achieving sufficiently strong and pure nitridation conditions. For previously reported ternary nitrides, high pressure ($\sim$8\,GPa) and/or strong nitriding agents such as \ch{NaN3} are required \cite{Cario2001,Choi2025,Talley2021,Klo2021,Weidemann2024,Zakutayev2024}.

Here, we report the discovery of a novel ternary nitride \ch{PrTaN2} achieved by thin film growth using molecular beam epitaxy (MBE). Compounds composed of lanthanides and tantalum have already been investigated in bulk form \cite{Cario2001,Choi2025,Weidemann2024}. However, this particular phase has not been reported to date. This suggests that thin film growth provides access to phases that are not readily obtained in bulk synthesis. Our MBE system offers several advantages in addressing the challenges mentioned above. E-beam evaporators are available in our chamber, which can handle high-melting-point elements. Furthermore, a radical source using pure N$_2$ provides favorable nitridation conditions. Compared with bulk growth using ammonia or azide, the effective nitrogen chemical potential can be higher in MBE because nitrogen is supplied in the form of reactive radical species. This provides a highly nitriding environment for film growth. These advantages enabled the discovery of the novel ternary nitride \ch{PrTaN2}.

To confirm that the observed crystalline phase corresponds to a new material, we performed detailed X-ray diffraction (XRD) measurements and scanning transmission electron microscopy (STEM). There are multiple candidate phases in the Pr-Ta-N system other than \ch{PrTaN2}. Detailed crystallographic analysis allows us to rule out these possible candidates. Furthermore, we determined the space group and atomic coordinates by developing a fitting procedure based on the structure factor. These findings open a new avenue for the exploration of novel ternary nitrides using thin film growth.

\section{Methods}
\ch{PrTaN2} thin films were grown on \ch{YAlO3} (YAO) (001)$_\mathrm{c}$ substrates. For clarity in describing the substrate azimuth, we use the pseudo-cubic Miller index notation with the subscript "c". Our custom-designed MBE system is equipped with e-beam evaporators for generating Pr and Ta fluxes. Controlling the flux using e-beam evaporators is challenging due to the high melting point of tantalum (3017\,$^\circ$C). Nevertheless, the flux was precisely controlled using electron impact emission spectroscopy (EIES) \cite{Yamamoto2013}. The growth rate was maintained at 0.025\,nm/s, and 20 min growth resulted in a film thickness of 30\,nm. For nitridation, pure N$_2$ gas (99.9999\%) was introduced into a radio-frequency (RF) radical source. To create a strong nitriding environment, a custom-designed atomic nitrogen source was operated at 13.56\,MHz with an RF power of 400\,W. The nitrogen flow rate was set to 2.0\,sccm, corresponding to a background nitrogen pressure of 1 $\times$ 10$^{-5}$\,Torr. The substrate temperature during growth was monitored using a radiation pyrometer (Japan Sensor), and the growth temperature was 800$\,^\circ$C. The supplied cation ratio (Pr:Ta) was calibrated to 1:1 using EIES and a quartz crystal microbalance equipped in the MBE chamber. The ratio of the resultant film was determined by inductively coupled plasma measurements.

For structural determination, X-ray diffraction (XRD) measurements were performed at room temperature. We used the SAGA Light Source (SAGA-LS) BL15 with a wavelength of 1.00\,\AA, for the 00$l$ and in-plane XRD measurements (Fig. \ref{fig:00l} and Fig. \ref{fig:sym}(h)). We also used a Bruker D8 with a Cu K$\alpha$ X-ray source (wavelength 1.54\,\AA) for other crystallographic plane measurements, including symmetric scans and reciprocal space maps (Fig. \ref{fig:sym}(a) - (g) and Fig. \ref{fig:rsm}). The X-ray beam from the SAGA-LS has a photon flux at least 100 times higher than that from the Cu source. This high flux is useful for detecting weak diffraction peaks from small crystalline volumes. On the other hand, in terms of intensity stability, the Cu source provides a constant X-ray intensity, whereas the intensity from the SAGA-LS decreases over time. Therefore, the Cu source was used for fitting the measured XRD peaks, since peak intensity is the fitting target. For consistency in the fitting analysis, the lattice parameters presented in this study were obtained from the data measured using the Bruker D8.

%\subsection{Outline}
\section{Experimental Results}
\subsection{Novel ternary nitride \ch{PrTaN2}}
We have synthesized and identified a novel ternary nitride \ch{PrTaN2}. The film resistivity is higher than 10$^3\, \Omega$m at room temperature, and paramagnetic behavior is observed (see Supporting Information). Our XRD study reveals that the crystal system is orthorhombic, with lattice constants of $a=$ 3.97 $\pm$ 0.08 \,\AA, $b=$ 4.00 $\pm$ 0.08\,\AA, and $c=$ 4.02 $\pm$ 0.01\,\AA. The \ch{PrTaN2} thin film is grown along the $c$ direction on a YAO (001)$_\mathrm{c}$ substrate.

A related ternary nitride is \ch{LaReN2} ($P4/mmm, a=b=$ 3.97\,\AA, $c=$ 3.56\,\AA)\cite{Klo2021}. In that system, the perovskite nitride \ch{LaReN3} decomposes at 500-620\,$^\circ$C and is topotactically reduced to \ch{LaReN2}. In contrast, \ch{PrTaN2} is not obtained via such a reduction process but is directly grown on the substrate. We did not observe the formation of perovskite \ch{PrTaN3} within the accessible growth conditions. In the case of \ch{PrTaN3}, the valences of Pr and Ta are expected to be +4 and +5, respectively, whereas both Pr and Ta are considered to be in the +3 state in \ch{PrTaN2}. The ionization energy of Pr$^{3+}$ is 38.9\,eV lower than that of Pr$^{4+}$\cite{NIST_ASD}, which favors the formation of the trivalent state\cite{Sarmiento-Prez2015}.

\begin{figure}[htbp]
  \centering
  \includegraphics[width=\linewidth]{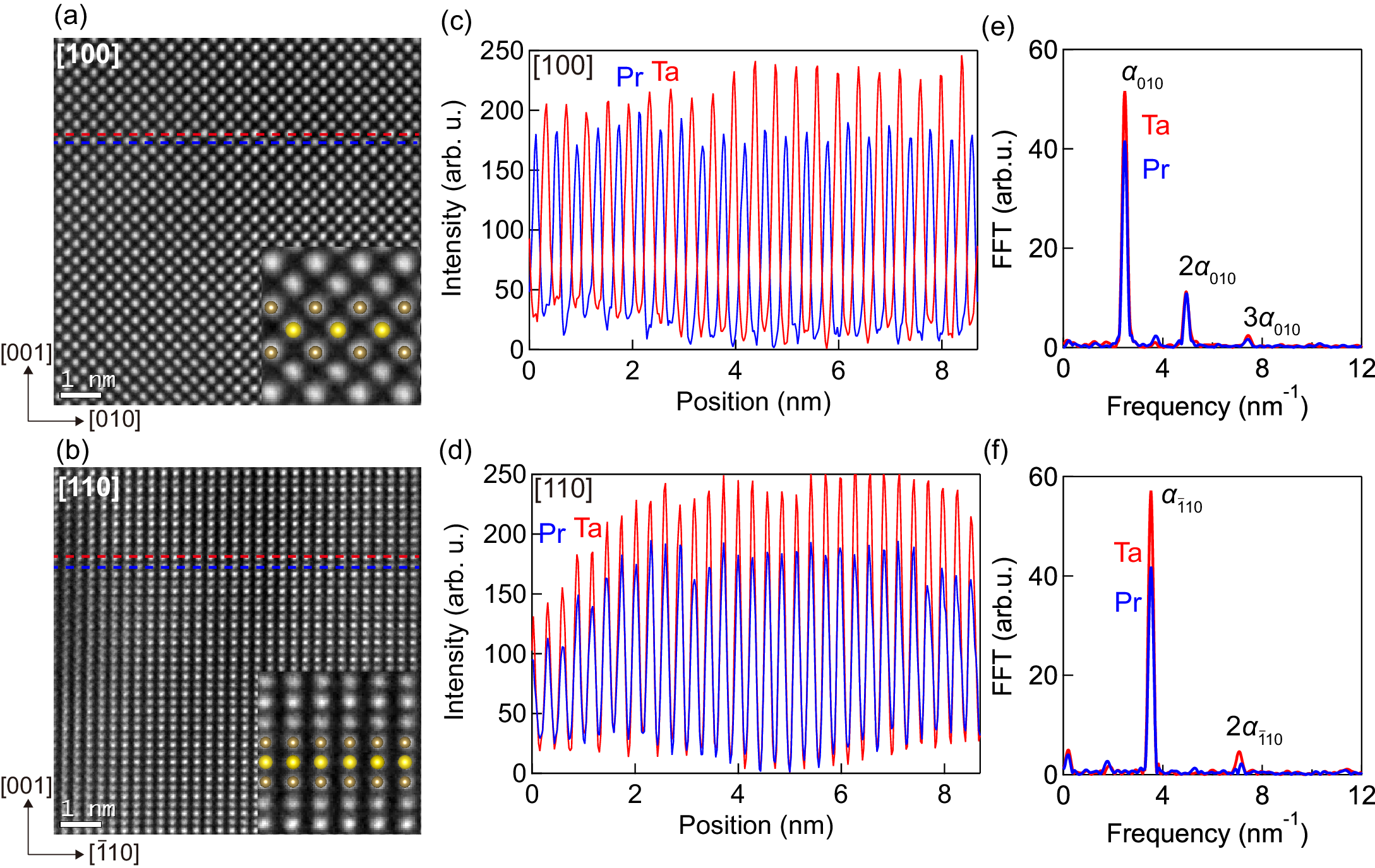}
  \caption{(a,b) High-angle annular dark field scanning transmission electron microscopy (HAADF-STEM) images in the [100] and [110] directions of \ch{PrTaN2}. The inset shows the expanded images with the schematic atomic images of \ch{PrTaN2}. The scale bar corresponds to 1\,nm. (c,d) Line intensity profiles of red and blue dashed lines in (a) and (b), respectively. Since the intensity of HAADF-STEM represents the atomic number, we can distinguish the cation positions, which means the red and blue lines represent Ta and Pr. For (c) and (d), Pr and Ta are located in alternate shifts in [001] direction. For the in-plane direction, while Pr and Ta are located half cycle change in (c), they are located same position in (d).(e,f) Fast Fourier transform (FFT) results of (c) and (d). The frequencies of the dominant peaks ($\alpha_{010}, \alpha_{\bar{1}10}$) agree with the corresponding $d$-spacing obtained by our XRD measurements. The harmonic peaks ($2\alpha_{010},3\alpha_{010}$ and $2\alpha_{\bar{1}10}$) are also observed.}
  \label{fig:stem}
\end{figure}

Figures \ref{fig:stem}(a) and (b) show high-angle annular dark-field scanning transmission electron microscopy (HAADF-STEM) images along the [100] and [110] directions of \ch{PrTaN2}. Clear atomic ordering is observed in both crystallographic directions. Since $I_{\rm{STEM}}\propto Z_a^{2.0}$\cite{Pennycook2012}, where $I_{\rm{STEM}}$ is the HAADF-STEM intensity and $Z_a$ is the atomic number, comparison of the line profiles allows identification of the atomic columns corresponding to Pr ($Z_a=59$) and Ta ($Z_a=73$). Figures \ref{fig:stem}(c) and (d) show the line profiles corresponding to the dashed lines in Figs. \ref{fig:stem}(a) and (b), respectively. The red curve exhibits higher intensity than the blue one, indicating Ta and Pr columns, respectively. 

In the HAADF-STEM image along the [100] direction (Fig. \ref{fig:stem}(a)), the in-plane positions of Pr and Ta are shifted by half a period. In contrast, in the [110] direction (Fig. \ref{fig:stem}(b)), Pr and Ta are located at the same in-plane positions.

Figures \ref{fig:stem}(e) and (f) are the fast Fourier transform (FFT) results of Figs. \ref{fig:stem}(c) and (d). The dominant peaks ($\alpha_{010}$ and $\alpha_{\bar{1}10}$) of Pr and Ta overlap, indicating that Pr and Ta share the same in-plane periodicity. The frequencies of $\alpha_{010}$ and $\alpha_{\bar{1}10}$ yield the corresponding $d$-spacings: $d_{010}=1/\alpha_{010} = 1/(2.54 \pm 0.14$\,nm$^{-1}) = 4.07 \pm 0.21$\,\AA, and $d_{110}=1/\alpha_{\bar{1}10} = 1/(3.52 \pm 0.13$\,nm$^{-1}) = 2.84 \pm 0.10$\,\AA. These $d$-spacings are in agreement with those obtained from our XRD measurements, as discussed later (see also Table \ref{tbl:dval}).

\subsection{Symmetric scans}

\begin{figure}[htbp]
  \centering
  \includegraphics[width=0.8\linewidth]{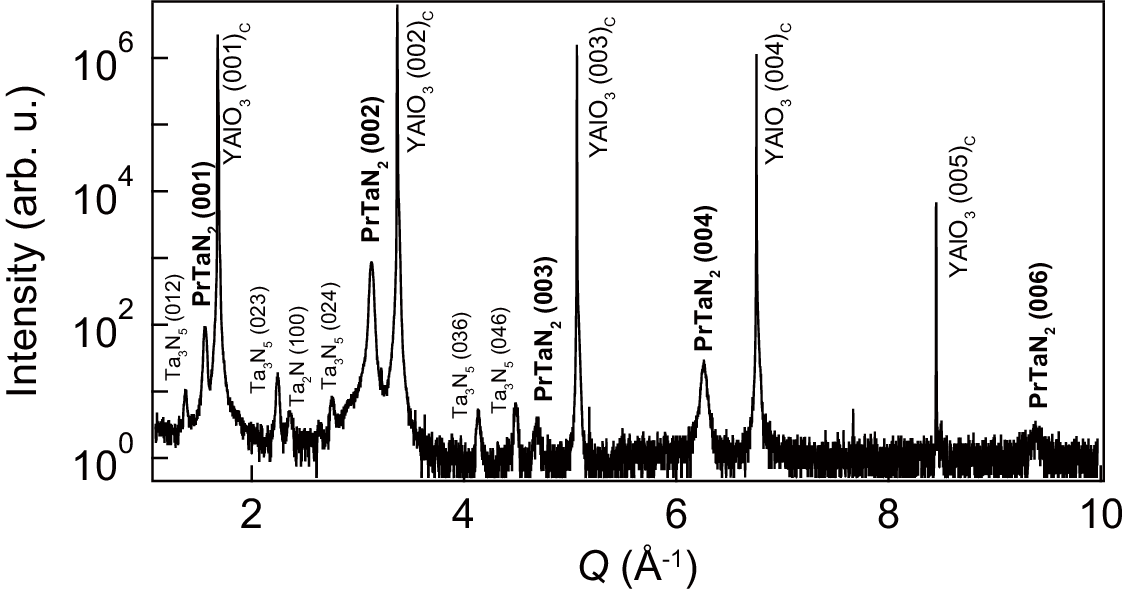}
  \caption{XRD symmetric scan in [001]$_\mathrm{c}$ direction of YAO measured in SAGA-LS with wave length of 1.00\,\AA. The crystalline coherency of \ch{PrTaN2} in [001] direction is confirmed with the higher order (00$l$) peaks up to (006). We also observed the Ta$_3$N$_5$ (012) and (023) peaks as well as their higher order diffraction peaks, which indicates that our RF radical source achieves the rich nitrogen environment. }
  \label{fig:00l}
\end{figure}

Detailed XRD measurements are key to confirming that the obtained phase corresponds to \ch{PrTaN2}. Figure \ref{fig:00l} shows the $\theta - 2\theta$ scan along the YAO [001]$_\mathrm{c}$ direction measured at the SAGA-LS with a wavelength of 1.00\,\AA. The \ch{PrTaN2} (001) peak and its higher-order diffractions up to (006) are observed, indicating high crystalline coherence along this direction. In addition, Ta$_3$N$_5$ (012), (023) and their higher order peaks are also observed as secondary phases.

Ta$_3$N$_5$ is metastable in the Ta-N system compared to TaN and Ta$_2$N. The presence of \ch{Ta3N5} as a secondary phase suggests that the growth was carried out under a high effective nitrogen chemical potential. Under such conditions, nitrogen incorporation is enhanced, which may facilitate the formation of nitrogen-rich phases such as \ch{PrTaN2} that are not readily obtained under bulk equilibrium synthesis conditions.

\begin{figure}[htbp]
  \centering
  \includegraphics[width=\linewidth]{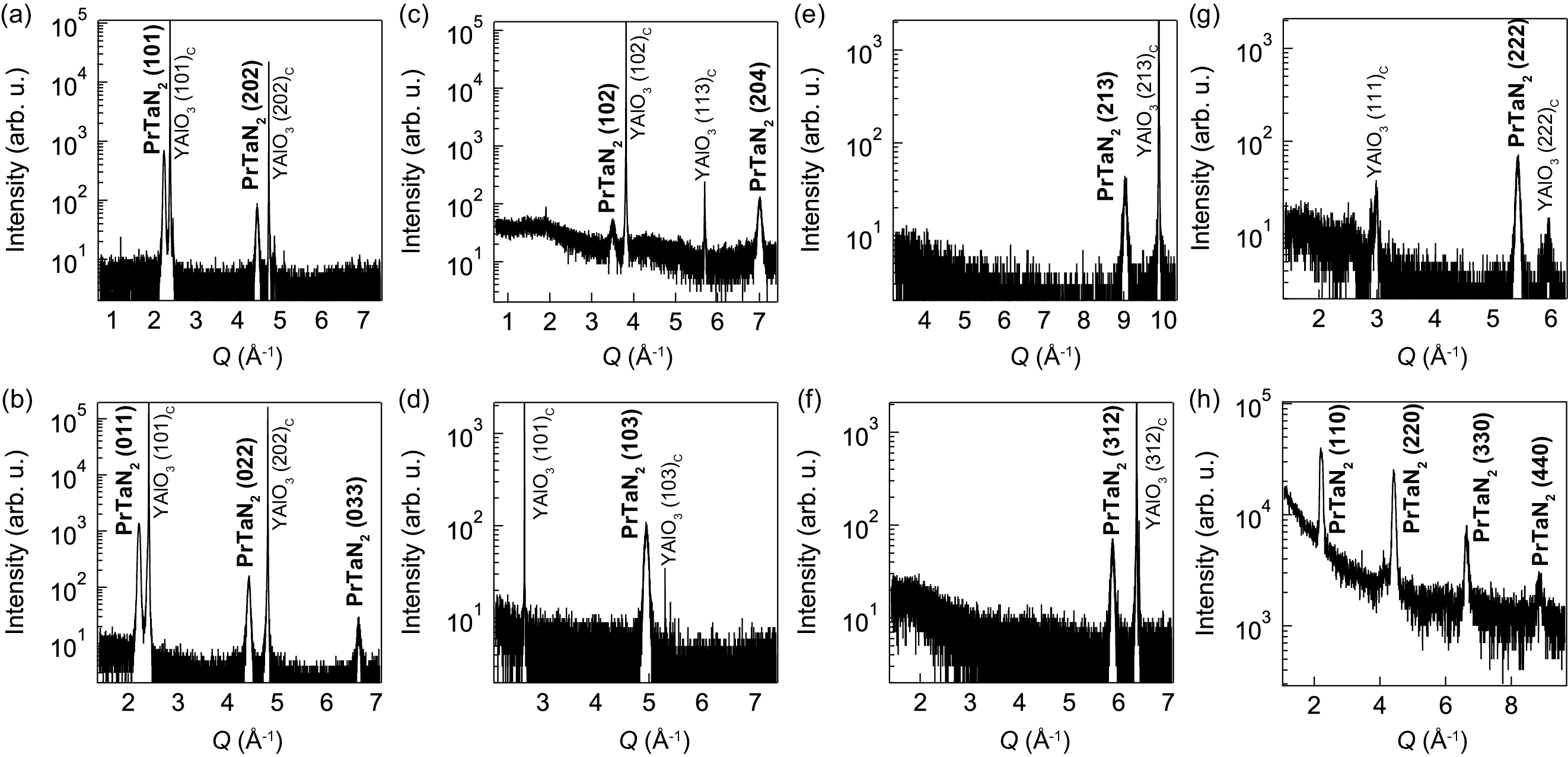}
  \caption{XRD symmetric scan aligned in the direction of (a) (101) ($\chi=45.5^\circ$), (b) (011) ($\chi=45.4^\circ$), (c) (102) ($\chi=63.0^\circ$), (d) (103) ($\chi=72.9^\circ$), (e) (213) ($\chi=37.0^\circ$), (f) (312) ($\chi=58.0^\circ$), and (g)(222) ($\chi=54.8^\circ$) \ch{PrTaN2} planes. (h) In-plane XRD measurement with $\chi=0.5^\circ$. The estimated $d$ spacing are presented in Table \ref{tbl:dval}.}
  \label{fig:sym}
\end{figure}

To investigate other crystallographic planes, we performed additional symmetric XRD scans with varying tilt angle $\chi$. Figures \ref{fig:sym}(a)-(g) show the results of symmetric scans corresponding to the (a) (101) ($\chi=45.5^\circ$), (b) (011) ($\chi=45.4^\circ$), (c) (102) ($\chi=63.0^\circ$), (d) (103) ($\chi=72.9^\circ$), (e) (213) ($\chi=37.0^\circ$), (f) (312) ($\chi=58.0^\circ$), and (g)(222) ($\chi=54.8^\circ$) \ch{PrTaN2} planes. Figure \ref{fig:sym}(h) shows the in-plane XRD results measured at SAGA-LS with the wave length of 1.00\,\AA. In Figs. \ref{fig:sym}(a) and (b), we measured the higher order peaks of (101) and (011) planes, which indicates the high crystalline coherency of our film. 

The $d$ spacing values in Figs. \ref{fig:sym}(a)-(d) are summarized in Table \ref{tbl:dval}. The $d$ values were determined using the Nelson-Riley fitting procedure \cite{Nelson1945} when higher-order diffractions were measured. The calculated $d$ values for an orthorhombic crystal system are given by the following relation:
\begin{eqnarray}
  \frac{1}{d} = \sqrt{\frac{h^2}{a^2}+\frac{k^2}{b^2}+\frac{l^2}{c^2}}.
\end{eqnarray}
For all measured crystallographic planes, the experimental $d$ values agree with the calculated ones within errors. Also, the $\varphi$ scans show the consistent $\varphi$ angles between neighboring peaks for (204) and (312) planes (see Supporting Information). These results are consistent with an orthorhombic crystal system, and the estimated lattice constants agree with the experimental data. 

In addition, we quantitatively evaluate the monoclinic distortion of the system, namely the deviation of $\beta$ from 90\,$^\circ$. The $d$-spacings for a monoclinic system are described as
\begin{eqnarray}
  \frac{1}{d} = \sqrt{\frac{1}{\sin^2 \beta} \left[\frac{h^2}{a^2}+\frac{k^2}{b^2}\sin^2 \beta+\frac{l^2}{c^2}-\frac{2hl\cos \beta}{ac}\right]}.
  \label{eq:mono}
\end{eqnarray}
Assuming that $\beta$ deviates from 90$^\circ$ within the uncertainty range of the experimentally determined $d$-spacings listed in Table \ref{tbl:dval}, Equation (\ref{eq:mono}) yields a range of 89.9$^\circ$$<\beta<$90.1$^\circ$. The deviation of $\beta$ from 90$^\circ$ is therefore sufficiently small. In other words, treating the system as orthorhombic does not lead to any inconsistency with the experimental results.

\begin{table}
  \caption{Experimental and calculated $d$ spacing values. When we measured one higher order peaks at least, $d$ is obtained by Nelson-Riley fitting.\cite{Nelson1945}}
  \label{tbl:dval}
  \centering
  \begin{tabular}{lll}
    \hline
    $hkl$  &  Exp. (\AA) & Calc. (\AA) \\
    \hline
    101 & 2.82 $\pm$ 0.03 & 2.82 $\pm$ 0.11\\
    011 & 2.83 $\pm$ 0.03 & 2.83 $\pm$ 0.11\\
    102 & 1.79 $\pm$ 0.01 & 1.79 $\pm$ 0.04\\
    103 & 1.27 $\pm$ 0.01 & 1.27 $\pm$ 0.03\\
    213 & 1.07 $\pm$ 0.004 & 1.07 $\pm$ 0.03\\
    312 & 1.07 $\pm$ 0.003 & 1.07 $\pm$ 0.05\\
    222 & 1.16 $\pm$ 0.004 & 1.15 $\pm$ 0.04\\
    110 & 2.84 $\pm$ 0.03 & 2.82 $\pm$ 0.14\\
    \hline
  \end{tabular}
\end{table}

\subsection{Epitaxial relationship and substrate choice}
\begin{figure}[htbp]
  \centering
  \includegraphics[width=\linewidth]{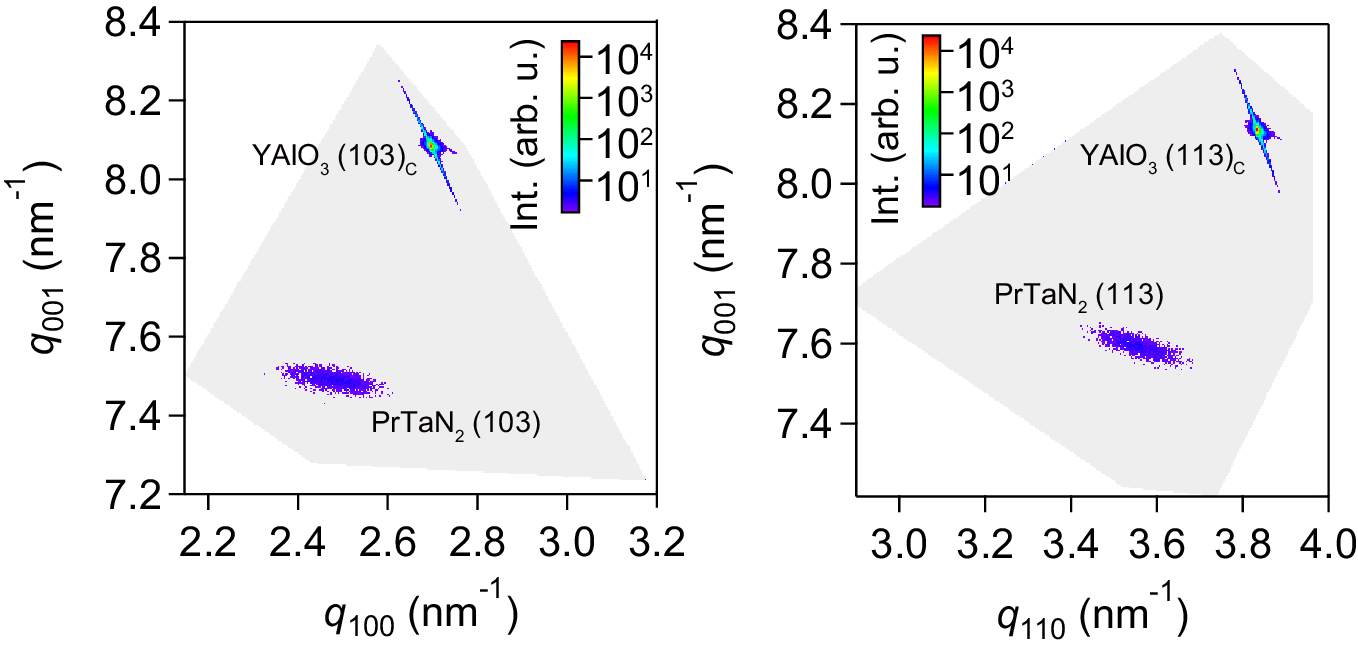}
  \caption{High resolution reciprocal space maps of the \ch{PrTaN2} thin film around the YAO (103)$_\mathrm{c}$ and YAO (113)$_\mathrm{c}$ peak. Logarithmic scale is used to express the intensity of the diffraction, which is shown as the color bar. The relaxed film is grown on the substrate, and the film and the substrate share the same crystalline axis.}
  \label{fig:rsm}
\end{figure}

The influence of the substrate is key to understanding the growth mechanism of \ch{PrTaN2}. Figures \ref{fig:rsm}(a) and (b) show reciprocal space maps around the YAO (103)$_\mathrm{c}$ and (113)$_\mathrm{c}$ diffractions. The reciprocal space coordinates $(q_x,q_z)$ for the diffraction peaks are summarized in Table \ref{tbl:rsm}. Here, $q_x$ corresponds to $q_{100}$ in Fig. \ref{fig:rsm}(a) and to $q_{110}$ in Fig. \ref{fig:rsm}(b), while $q_z$ corresponds to $q_{001}$.

The ratios $q_z/q_x$ for the (103) and (113) reflections are nearly identical for the \ch{PrTaN2} film and the YAO substrate, indicating that the film is essentially strain-free. Thus, the lattice constants are not significantly affected by substrate-induced strain. On the other hand, the in-plane crystal orientation is not completely arbitrary but reflects the four-fold symmetry of the YAO substrate. This behavior suggests preferentially oriented growth influenced by the substrate rather than fully coherent epitaxy.

The choice of substrate is an important growth parameter for obtaining ternary nitride thin films. One difficulty in substrate selection is the lack of commercially available nitride substrates with cubic or tetragonal symmetry. Maintaining a high-quality interface between dissimilar materials during crystal growth is generally challenging. So far, a variety of oxide substrates have helped address this issue.

As seen in the RSM measurements, a coherent epitaxial relationship between \ch{PrTaN2} and the YAO substrate is not established. The pseudo-cubic lattice constant of YAO is 3.71\,\AA\, which does not match any characteristic atomic spacing in \ch{PrTaN2}. Oxide substrates such as \ch{SrTiO3} and \ch{KTaO3} offer a range of lattice constants around 3.9\,\AA-4.0\,\AA, which closely match the $a$, $b$, and $c$ lattice constants of \ch{PrTaN2}. However, these substrates are not suitable for growth under the present conditions because their surfaces are not sufficiently resistant to nitridation. The high chemical potential associated with nitrogen radicals and/or ionic nitrogen generated by the RF radical source can degrade the substrate surface by inducing oxygen deficiency.

\ch{YAlO3} has a higher oxygen vacancy formation energy ($\sim$ 22\,eV)\cite{Fu2018} compared to \ch{SrTiO3} ($\sim$5.9\,eV)\cite{Rusevich2021} and \ch{KTaO3} ($\sim$5.3\,eV)\cite{Choi2011}. This suggests that \ch{YAlO3} is more robust against reduction under highly nitriding conditions. Therefore, \ch{YAlO3} is one of the most suitable choices for ternary nitride thin film growth.

\begin{table}[htbp]
  \caption{Reciprocal space coordinates observed in Fig. \ref{fig:rsm}. $q_x$ corresponds to $q_{100}$ in Fig. \ref{fig:rsm}(a) and to $q_{110}$ in Fig. \ref{fig:rsm}(b), while $q_z$ corresponds to $q_{001}$.}
  \label{tbl:rsm}
  \centering
  \begin{tabular}{lllll}
    \hline
    $hkl$ & Material  &  $q_{z}$ (nm$^{-1}$) & $q_{x}$ (nm$^{-1}$) & $q_{z}/q_{x}$ \\
    \hline
    \multirow{2}{*}{103}&\ch{PrTaN2} & 7.49 $\pm$ 0.08 & 2.48 $\pm$ 0.21 & 3.02 $\pm$ 0.27 \\
    &YAlO$_3$ & 8.09 $\pm$ 0.01 & 2.69 $\pm$ 0.01 & 3.00 $\pm$ 0.01 \\
    \multirow{2}{*}{113}&\ch{PrTaN2} & 7.59 $\pm$ 0.07 & 3.56 $\pm$ 0.16 & 2.12 $\pm$ 0.10\\
    &YAlO$_3$ & 8.13 $\pm$ 0.01 & 3.83 $\pm$ 0.01& 2.12 $\pm$ 0.01 \\
    \hline
  \end{tabular}
\end{table}

\section{Discussion}
\subsection{Determination of chemical formula and crystal structure}

In exploring new materials in thin film form, careful determination of the chemical formula and crystal structure is essential. Even when the observed XRD peaks appear consistent with a given structure, there remains a degree of freedom in assigning the $hkl$ indices in terms of integer multiples. In other words, the choice of the unit cell based solely on XRD data is not necessarily unique. To constrain such ambiguity and determine a specific chemical formula and crystal structure, it is necessary to analyze diffraction peaks from multiple crystallographic planes in combination with STEM observations. As mentioned in the Introduction, various ternary nitrides composed of lanthanides and tantalum have been reported. In the present study, we systematically rule out alternative candidate phases based on the combined XRD and STEM analyses.

\ch{Pr3Ta2N6} is a Ruddlesden-Popper related phase with tetragonal symmetry ($I4/mmm$) with $a = b = 4.06$\,\AA\, and $c = 19.6$\,\AA\cite{Cario2001}. According to the extinction rules for the $I4/mmm$ space group, no diffraction peak is expected in the measured reciprocal space around the (113) reflection This is inconsistent with the experimental observation shown in Fig. \ref{fig:rsm}(b).

Perovskite (\ch{PrTaN3}) and the Ruddlesden--Popper phase with $n=2$ (\ch{Pr2TaN4}) are also possible candidate structures. \ch{CeTaN3}\cite{Choi2025} and \ch{Ce2TaN4}\cite{Weidemann2024} are reported analogous compounds. However, these possibilities are unlikely based on the observed trivalent state of Pr. Considering that the maximum valence of Ta is +5, Pr is in a tetravalent state in these structures. We also grew isostructural \ch{NdTaN2} on the YAO substrate (see Supporting Information.) Since Nd and Ta are trivalent, the formation of phases such as \ch{PrTaN3} and \ch{Pr2TaN4} is unlikely. 

Another possible candidate is the brownmillerite phase \ch{Pr2Ta2N5} If our film were \ch{Pr2Ta2N5}, the lattice constants would be estimated as $a=$ 5.64\,\AA, $b=$ 5.66\,\AA, and $c=$ 16.1\,\AA, based on our XRD results. These lattice parameters were obtained by indexing the XRD peaks so that each parameter takes reasonable values compared to those of known brownmillerite compounds. In this scenario, the diffraction peaks observed in the $00l$ scan (Fig. \ref{fig:00l}) can be assigned to the 004 diffraction and its higher-order diffractions (008, 0012,\dots). The space groups consistent with the extinction rules are $F2dd$ and $Fddd$. If these space groups were valid, the XRD peak observed in the reciprocal space mapping around YAO(113)$_\mathrm{c}$ (Fig. \ref{fig:rsm}(b)) could be indexed as (1 12 1). However, this reflection is forbidden by the extinction rules of both $F2dd$ and $Fddd$, indicating that the brownmillerite phase is unlikely.

\ch{Pr2TaN3} with orthorhombic symmetry is also a possible candidate. As a similar example, \ch{Ce2TaN3} has been reported\cite{Broll1995}. However, this possibility can be ruled out based on the HAADF-STEM measurements. We performed STEM image simulations for \ch{Pr2TaN3} using ReciPro\cite{Seto2022} (see Supporting Information). In the [011] direction, the experimental data show an atomic sequence of $\cdots -$Pr$-$Ta$-$Pr$-$Ta$- \cdots$. In contrast, the simulation exhibits a sequence of $\cdots -$Pr$-$Pr$-$Ta$-$Pr$-$Pr$- \cdots$. This discrepancy is inconsistent with the experimental observations, indicating that the \ch{Pr2TaN3} structure is unlikely.

\ch{ScTaN2} has a hexagonal structure ($P6_3/mmc$) and features trivalent Ta, making it a relevant reference polymorph when considering possible crystal structures of \ch{PrTaN2}. However, the possibility of a hexagonal structure in the present case is unlikely. Our $\varphi$ scans show no indication of hexagonal symmetry (see Supporting Information). The $\varphi$ scans of the (204) and (312) reflections exhibit fourfold and eightfold peak patterns, respectively. This indicates that the crystal system does not possess sixfold rotational symmetry.

\subsection{Determination of space group and atomic coordinates}
Determining the space group and atomic coordinates is a key challenge in investigating a new material in thin-film form. In general, powder diffraction patterns allow determination of the space group and atomic coordinates through Rietveld analysis. The XRD $\theta$--$2\theta$ scan provides sufficient crystallographic information to determine the crystal structure in powder samples.

In contrast, for single-crystalline thin films, diffraction peaks are distributed in reciprocal space, and only a limited number of crystallographic planes can be measured. Although extinction rules help narrow down possible space groups, they do not always uniquely determine the space group. For example, $P222$ and $Pmmm$, which belong to the same Laue class, cannot be distinguished solely by extinction rules \cite{Looijenga-Vos2006}.

In principle, determining the space group and crystal structure is equivalent to solving the structure factor equations,
\begin{eqnarray}
F_{hkl}=\sum_\nu f_\nu\exp[2\pi i(hx_\nu+ky_\nu+lz_\nu)],
\label{eq:F}
\end{eqnarray}
where $f_\nu$ is the scattering factor, and $\nu$ represents atomic species. In the case of \ch{PrTaN2} with number of formula units per unit cell $Z=1$, $\nu$ corresponds to Pr, Ta, and two inequivalent nitrogen atoms (denoted as N1 and N2). In this case, one needs to solve the equations with ($3\times4=$) 12 parameters for the atomic positions, which is generally impractical to determine directly. 

\begin{figure}[htbp]
  \centering
  \includegraphics[width=\linewidth]{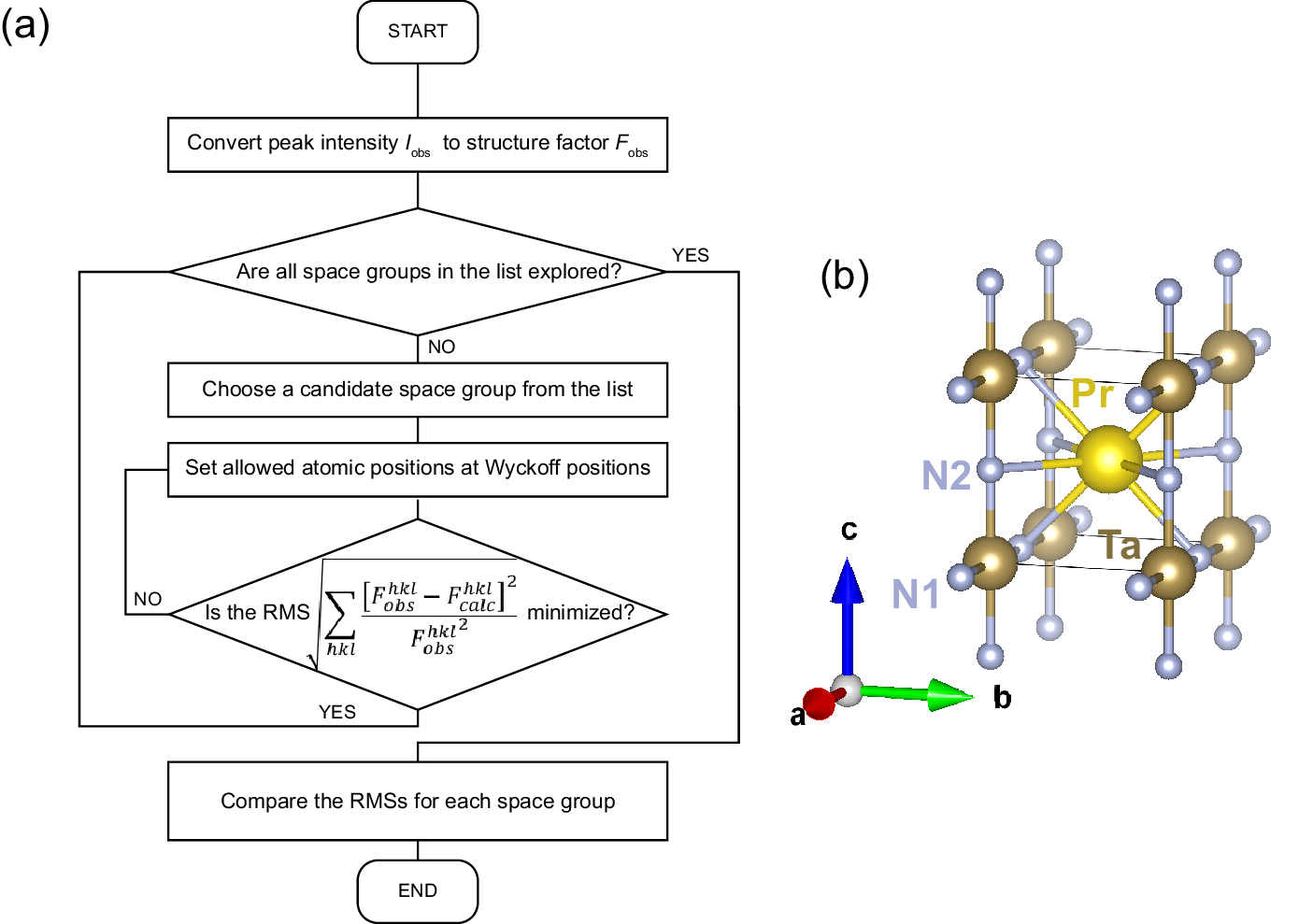}
  \caption{(a)Fitting procedure of eq. (\ref{eq:F}) to the experimental data. To convert the peak intensity $I_{\rm{obs}}$ to the structure factor $F_{\rm{obs}}$, we use eq. (\ref{eq:I-F}). By using this procedure, we can decrease the number of fitting parameter using Wyckoff positions. (b) Crystal structure of \ch{PrTaN2} reproduced by the estimated atomic coordinates.}
  \label{fig:sim}
\end{figure}

Here, we present a fitting procedure that enables determination of the crystal structure even from thin-film XRD data (Fig. \ref{fig:sim}(a)): (i) Convert the observed XRD peak intensities $I_{\mathrm{obs}}$ into structure factors $F_{\mathrm{obs}}$. (ii) Select candidate space groups based on extinction rules. (iii) Assign allowed atomic positions according to Wyckoff positions. (iv) Solve an optimization problem to minimize the root-mean-square (RMS) residual between experimental and calculated structure factor values for the observed $hkl$ reflections. (v) Compare the RMS values among different space groups and make a decision after screening by chemical bond length. We note that we assumed the smallest translational periodicity compatible with the observed diffraction pattern and adopted a $Z = 1$ orthorhombic average structure. The reasons why this assumption is justified are discussed later in this section.

For step (i), one needs to account for extrinsic factors that affect the peak intensity. The relationship between $I_{\mathrm{obs}}$ and $F_{\mathrm{obs}}$ is described as \cite{Birkholz2005}
\begin{eqnarray}
I^{hkl}_{\rm{obs}}(\theta) \propto C^2(\theta)L(\theta)A_{\theta 2 \theta}(\theta) |F^{hkl}_{\rm{obs}}|^2,
\label{eq:I-F}
\end{eqnarray}
Note that multiplicity and geometrical factors can be neglected in the case of thin films. The polarization factor $C^2(\theta)= (1+\cos^2\theta)/2$ represents the angular dependence of the scattering intensity arising from Thomson and Compton scattering. The Lorentz factor $L(\theta) = (\sin(2\theta))^{-1}$ accounts for the geometrical effect associated with the measurement configuration. The absorption factor $A_{\theta 2\theta}(\theta)= 1 - \exp[-\mu t/(\sin\theta\cos\chi)]$ describes the attenuation of X-rays in the film, where $\mu$ is the linear attenuation coefficient (= $1.96\times 10^{5}$\,/m)\cite{NIST_MAC}, $t$ is the film thickness (= 30 $\pm$ 0.1\,nm), and $\chi$ is the tilt angle of the goniometer.

For steps (ii) and (iii), the extinction rules indicate that the candidate orthorhombic space groups are $P222$ (16), $P22_12$ (17), $Pmm2$ (25), and $Pmmm$ (47). In the fitting procedure, one space group is selected from this list, and atomic positions are assigned according to the corresponding Wyckoff positions. The Wyckoff orbits ensure that the symmetry of each hypothetical structure is consistent with the selected space group.

For step (iv), we solved an optimization problem for the atomic positions to minimize the RMS. As an example, in the case of $Pm2m$, the Wyckoff positions can be assigned as Pr (0, $y_{\rm{Pr}}$, 1/2), Ta (1/2, 1/4, 0), N1 (0, $y_{\rm{N1}}$, 0), and N2 (1/2, $y_{\rm{N2}}$, 1/2). In this case, eq.(\ref{eq:F}) is simplified to include three free fitting parameters ($y_{\rm{Pr}}, y_{\rm{N1}}, y_{\rm{N2}}$). In other words, this procedure reduces the number of fitting parameters by utilizing the constraints imposed by Wyckoff positions. When there are no free variables, as in Wyckoff letter 1a (0,0,0) of $P222$, an atom is simply placed at that position, and no fitting is performed for that site. We note that it is unlikely that atoms occupy general positions. For example, in the case of $Pmmm$, the multiplicity of the general position is eight (Wyckoff letter 8A). This is physically unlikely because it cannot maintain the stoichiometric ratio of \ch{PrTaN2}, and it would also lead to unrealistic bond lengths (typically < 1.0\,\AA). 

The positions of Pr and Ta were taken from the STEM images. This procedure further reduces the number of free variables, thereby making the statistical errors in the fitting results sufficiently small. The cation positions determined from STEM were Pr(0.51 $\pm$ 0.04, 0.53 $\pm$ 0.04, 0.49 $\pm$ 0.01) relative to Ta(0,0,0). Within the uncertainty of the error bars, the only atomic position that Pr could occupy while maintaining the stoichiometry of \ch{PrTaN2} and satisfying the constraints imposed by the allowed Wyckoff positions was (0.5,0.5,0.5). Therefore, the actual fitting was performed using Ta(0,0,0) and Pr(0.5,0.5,0.5).

For step (v), inspection of the RMS values obtained from the fitting revealed that it is difficult to uniquely determine a single structural model solely on the basis of the RMS values, even among the best candidates. Comparing the top five results that correspond to structurally distinguishable candidates, the RMS values ranged only from 0.1735 to 0.1750, indicating that such small differences are insufficient for definitive structure determination. We note that the worst RMS is 44.5 on $Pmmm$. Therefore, additional screening was performed based on the Pr-N bond distances.

Among the candidates exhibiting nearly optimal RMS values, the nitrogen positions relative to Pr(0.5,0.5,0.5) can be broadly classified into the following categories:(1) sharing the $a$-axis (i.e., having the same $b$ and $c$ coordinates), (2) sharing the $b$-axis, (3) sharing the $c$-axis, and (4) sharing none of the axes, with N located along the in-plane diagonal directions (e.g., N(0.5,0,0) or N(0,0.5,0)).

For cases (1)-(3), considering that the lattice constants $a$, $b$, and $c$ are all approximately 4.0\,\AA, the resulting Pr-N distances are about 2.0\,\AA. To the best of our knowledge, no known compound exhibits such a short Pr-N bond distance. The simple sum of the ionic radii of Pr and N is approximately 2.5-2.8\,\AA, and the Pr-N distance in rock-salt PrN is 2.58\,\AA. For these reasons, only category (4) is considered physically plausible. In this case, the Pr-N distance is approximately 4.0\,\AA/$\sqrt{2}$ = 2.8\,\AA. The space group is estimated to be $P222$ with best RMS = 0.174.  The atomic coordinates and the bond lengths are shown in Table \ref{tbl:fit} and Table \ref{tbl:bnd}. Figure \ref{fig:sim}(b) shows the crystal structure of \ch{PrTaN2} using the estimated atomic coordinates.

The chosen Wyckoff positions for nitrogen are 1b and 1d, which do not have any free fitting parameter. The obtained structure shows some similarity to \ch{LaReN2}; however, a notable difference is that the lattice constants $a$, $b$, and $c$ in \ch{PrTaN2} are nearly identical. Although structural anisotropy is observed along the $b$-axis in \ch{PrTaN2}, no corresponding shortening of only the $b$-axis lattice constant is found. 

The coordination number for Pr and Ta are eight and four, respectively. In ternary nitrides containing Ta, the coordination number of Ta tends to be six, reflecting the Ta high oxidation state (+5) \cite{Niewa1998}. Examples in which Ta has a coordination number of four are rare. \ch{CsTaN2}, in which Ta forms \ch{TaN4} tetrahedra, is the only known example\cite{Cordes2019}. In this context, the possibility that excess nitrogen, lacking long-range order, is incorporated into interstitial sites near Ta cannot be excluded. If the presence of such interstitial nitrogen between the Ta-N planes is assumed, the larger interlayer spacing in \ch{PrTaN2} compared with \ch{LaReN2}\cite{Klo2021} may be understood (\ch{PrTaN2}: 3.97\,\AA; \ch{LaReN2}: 3.56\,\AA).Similar phenomena, in which excess anions occupying interstitial sites lead to an expansion of the interlayer spacing, are observed in layered oxide compounds \cite{Adachi1992,Mercey1995,Tomaschko2012}. A similar mechanism may therefore be operative in \ch{PrTaN2}.

We also tested lower-symmetry models such as monoclinic $P2$, where nitrogen atoms are no longer constrained to symmetry-equivalent positions and can move more independently. Such models effectively introduce additional crystallographically distinct nitrogen environments while keeping the same translational cell and $Z=1$. However, even with this increased flexibility, the structure-factor fit was not improved significantly (RMS = 0.173). Therefore, the XRD data do not support the need for additional independent nitrogen sites beyond those already present in the higher-symmetry model. Together with the absence of clear superlattice reflections, this indicates that the higher-symmetry $Z=1$ average structure is sufficient to describe the observed diffraction data.

\begin{table}[htbp]
\centering
\begin{tabular}{c c l c c c c}
\hline
Label & Element & Wyckoff position & $x$ & $y$ & $z$ \\
\hline
Pr & Pr & 1h (1/2, 1/2, 1/2) & 0.5 & 0.5 & 0.5 \\
Ta & Ta & 1a (0, 0, 0) & 0.0 & 0.0 & 0.0 \\
N1 & N & 1b (1/2, 0, 0) & 0.5 & 0.0 & 0.0 \\
N2 & N & 1d (0, 0, 1/2) & 0.0 & 0.0 & 0.5 \\
\hline
\end{tabular}
\caption{Optimized atomic coordinates and Wyckoff positions by our fitting procedure (Fig. \ref{fig:sim}(a)). According to our STEM measurements, the cation positions were Pr(0.51 $\pm$ 0.04, 0.53 $\pm$ 0.04, 0.49 $\pm$ 0.01) relative to Ta(0,0,0). Among the candidate space groups identified from the extinction rules ($P222$ (16), $P22_12$ (17), $Pmm2$ (25), and $Pmmm$ (47)), the Pr position within this error range is Pr(0.5,0.5,0.5).}
\label{tbl:fit}
\end{table}

\begin{table}[htbp]
\centering
\begin{tabular}{c c l c c c c}
\hline
Element & Bond length (\AA)\\
\hline
Pr - N1 & 2.85  \\
Pr - N2 & 2.84  \\
Ta - N1 & 1.95  \\
Ta - N2 & 1.97  \\
\hline
\end{tabular}
\caption{The chemical bond length between cations (Pr and Ta) and nitrogens (N1 and N2).}
\label{tbl:bnd}
\end{table}

\subsection{Structure factor analysis of \ch{PrTaN2}}
\subsubsection{Symmetry Operations and General Coordinates}
In this section, we examine in detail the structure factors calculated from the space group obtained by the fitting procedure. The $P222$ space group has primitive lattice centering and three mutually perpendicular twofold rotation axes ($2_x, 2_y, 2_z$) parallel to the $a, b,$ and $c$ axes, intersecting at the origin $(0,0,0)$.

The general positions generated by these operations for an atom at a general coordinate $(x, y, z)$ are:

1. $(x, y, z)$ — Identity ($x, y, z$)\par
2. $(\bar{x}, \bar{y}, z)$ — 2-fold axis along $z$: rotation by $180^\circ$ around $(0,0,z)$\par
3. $(\bar{x}, y, \bar{z})$ — 2-fold axis along $y$: rotation by $180^\circ$ around $(0,y,0)$\par
4. $(x, \bar{y}, \bar{z})$ — 2-fold axis along $x$: rotation by $180^\circ$ around $(x,0,0)$\par

\subsubsection{General Structure Factor Formula}

For an atom sitting in a general position $(x,y,z)$, its contribution must be summed over all 4 symmetrically equivalent positions generated above. Let us substitute these 4 coordinates into the exponential sum using eq. (\ref{eq:F}):
\begin{eqnarray}
F_{hkl} = f \left[ e^{2\pi i (hx + ky + lz)} + e^{2\pi i (-hx - ky + lz)} + e^{2\pi i (-hx + ky - lz)} + e^{2\pi i (hx - ky - lz)} \right]
\end{eqnarray}
Using grouping terms strategically, we can simplify this expression. Now, we group the terms to separate the $z$ and $-z$ dependencies:
\begin{eqnarray}
F_{hkl} = f \left[ e^{2\pi i lz} \left( e^{2\pi i(hx+ky)} + e^{-2\pi i(hx+ky)} \right) + e^{-2\pi i lz} \left( e^{2\pi i(-hx+ky)} + e^{-2\pi i(-hx+ky)} \right) \right]
\end{eqnarray}
Using Euler's formula ($\exp(i\theta) = \cos\theta + i\sin\theta$) and the identity $e^{i\theta} + e^{-i\theta} = 2\cos\theta$:
\begin{eqnarray}
F_{hkl} = 2f \left[ e^{2\pi i lz} \cos(2\pi(hx+ky)) + e^{-2\pi i lz} \cos(2\pi(-hx+ky)) \right]
\end{eqnarray}
Since $\cos(-\alpha+\beta) = \cos(\alpha-\beta)$, and expanding the outer exponentials into cosines and sines ($e^{\pm i\theta} = \cos\theta \pm i\sin\theta$), this simplifies to:
\begin{eqnarray}
F_{hkl} = 4f \left[ \cos(2\pi hx)\cos(2\pi ky)\cos(2\pi lz) - i \sin(2\pi hx)\sin(2\pi ky)\sin(2\pi lz) \right]
\end{eqnarray}

\subsubsection{Real and Imaginary Parts}

Unlike centrosymmetric space groups (like $Pmmm$) where the imaginary part vanishes because every atom at $(x,y,z)$ has an inversion counterpart at $(\bar{x},\bar{y},\bar{z})$, $P222$ is non-centrosymmetric. Therefore, the structure factor is complex and contains both a real part ($A$) and an imaginary part ($B$):
\begin{eqnarray}
F_{hkl} = A + iB
\end{eqnarray}
For a general reflection:
\begin{eqnarray}
A &=& 4 \sum_{\nu} f_\nu \cos(2\pi hx_\nu)\cos(2\pi ky_\nu)\cos(2\pi lz_\nu)\\
B &=& -4 \sum_{\nu} f_\nu \sin(2\pi hx_\nu)\sin(2\pi ky_\nu)\sin(2\pi lz_\nu)
\end{eqnarray}

\subsubsection{Special Reflections (Systematic Dynamic Absences)}
Because $P222$ is a primitive lattice without glide planes or screw axes, there are no systematic absences (extinctions) for general $(hkl)$ reflections caused by translational symmetry elements. However, specific projections yield special constraints on the imaginary component:

For $(h00)$, $(0k0)$, and $(00l)$ reflections: If two of the Miller indices are zero, the imaginary part $B$ drops to 0 because $\sin(0) = 0$. For instance, for $(h00)$:
\begin{eqnarray}
A = 4 \sum_{\nu} f_\nu \cos(2\pi hx_\nu), \quad B = 0
\end{eqnarray}

\subsubsection{Application to Wyckoff Positions (Special Positions)}

When atoms sit on special positions, their coordinates simplify significantly, which reduces the number of parameters needed during structure factor fitting (see Table \ref{tbl:wyc_N}). If ones incorporate this into the thin-film intensity profile fitting code, using these explicit geometric simplifications for $A$ and $B$ instead of computing a raw exponential matrix significantly accelerates the refinement loop.

To calculate the structure factors explicitly for the $(111)$, $(222)$, and $(333)$ reflections of the \ch{PrTaN2} thin film in the $P222$ space group, the atomic coordinates are :

- $\text{Ta}$ is fixed at the origin: $(0, 0, 0)$.\par
- $\text{Pr}$ is fixed at the body center: $(\frac{1}{2}, \frac{1}{2}, \frac{1}{2})$.\par
- $\text{N}$ atoms occupy the category (4) in-plane diagonal positions, putting them at the edge-centers: $(\frac{1}{2}, 0, 0)$ and $(0, 0, \frac{1}{2})$.\par

For these specific special coordinates, the imaginary components $B$ vanish completely because all values contain a $\sin(0)$ or $\sin(\pi) = 0$. The general formula simplifies down to:
\begin{eqnarray}
F_{hkl} = \sum_{\nu} f_\nu \cos[2\pi(hx_\nu + ky_\nu + lz_\nu)]
\end{eqnarray}

\paragraph{Calculation for the $(111)$ Peak}

Substitute the atomic positions into the simplified phase summation:
\begin{eqnarray}
F_{111} &= f_{\text{Ta}}\cos[2\pi(0)] + f_{\text{Pr}}\cos\left[2\pi\left(\frac{1}{2} + \frac{1}{2} + \frac{1}{2}\right)\right] + f_{\text{N1}}\cos\left[2\pi\left(\frac{1}{2}\right)\right] + f_{\text{N2}}\cos\left[2\pi\left(\frac{1}{2}\right)\right] \\
&= f_{\text{Ta}}\cos(0) + f_{\text{Pr}}\cos(3\pi) + f_{\text{N1}}\cos(\pi) + f_{\text{N2}}\cos(\pi)
\end{eqnarray}

Using $\cos(0) = 1$ and $\cos(\pi) = \cos(3\pi) = -1$:

\begin{eqnarray}
F_{111} = f_{\text{Ta}} - f_{\text{Pr}} - 2f_{\text{N}}
\label{eq:111}
\end{eqnarray}
Since $f_{\text{Ta}} (Z_a=73)$ and $f_{\text{Pr}} (Z_a=59)$ partially cancel each other out, and the nitrogen atoms scatter completely out of phase, $|F_{111}|$ drops drastically toward zero. This explicitly explains why the $(111)$ peak is systematically absent/unobserved in Fig. \ref{fig:sym}(g).

\paragraph{Calculation for the $(222)$ Peak}

Substitute the atomic positions for $h=2, k=2, l=2$:

\begin{eqnarray}
F_{222} &= f_{\text{Ta}}\cos[2\pi(0)] + f_{\text{Pr}}\cos\left[2\pi\left(\frac{2}{2} + \frac{2}{2} + \frac{2}{2}\right)\right] + f_{\text{N1}}\cos\left[2\pi\left(\frac{2}{2}\right)\right] + f_{\text{N2}}\cos\left[2\pi\left(\frac{2}{2}\right)\right] \\
&= f_{\text{Ta}}\cos(0) + f_{\text{Pr}}\cos(6\pi) + f_{\text{N1}}\cos(2\pi) + f_{\text{N2}}\cos(2\pi)
\label{eq:222}
\end{eqnarray}

Using $\cos(0) = \cos(2\pi) = \cos(6\pi) = 1$:

\begin{eqnarray}
F_{222} = f_{\text{Ta}} + f_{\text{Pr}} + 2f_{\text{N}}
\end{eqnarray}
Every single atomic sublattice scatters exactly in phase. This constructive interference yields a massive structure factor magnitude, making the $(222)$ peak highly visible and prominent in Fig. 3(g).

\paragraph{Calculation for the $(102)$ Peak}

Substitute the atomic positions for $h=1, k=0, l=2$:

\begin{eqnarray}
F_{102} &= f_{\text{Ta}}\cos[2\pi(0)] + f_{\text{Pr}}\cos\left[2\pi\left(\frac{1}{2} + \frac{0}{2} + \frac{2}{2}\right)\right] + f_{\text{N1}}\cos\left[2\pi\left(\frac{1}{2}\right)\right] + f_{\text{N2}}\cos\left[2\pi\left(\frac{2}{2}\right)\right]
\label{eq:102} \\
&= f_{\text{Ta}}\cos(0) + f_{\text{Pr}}\cos(3\pi) + f_{\text{N1}}\cos(\pi) + f_{\text{N2}}\cos(2\pi)
\end{eqnarray}

Using $\cos(0) = \cos(2\pi) = 1$ and $\cos(\pi) = \cos(3\pi) = -1$:

\begin{eqnarray}
F_{102} = f_{\text{Ta}} - f_{\text{Pr}}
\end{eqnarray}
Since the contributions from N1 and N2 completely cancel each other, the structure factor of (102) is larger than that of (111) ($|F_{102}| > |F_{111}|$). As shown in Fig. \ref{fig:sym}(c), we observed the (102) peak, in contrast to the absence of the (111) peak in Fig. \ref{fig:sym}(g).

\paragraph{Calculation for the $(204)$ Peak}

Substitute the atomic positions for $h=2, k=0, l=4$:

\begin{eqnarray}
F_{204} &= f_{\text{Ta}}\cos[2\pi(0)] + f_{\text{Pr}}\cos\left[2\pi\left(\frac{2}{2} + \frac{0}{2} + \frac{4}{2}\right)\right] + f_{\text{N1}}\cos\left[2\pi\left(\frac{2}{2}\right)\right] + f_{\text{N2}}\cos\left[2\pi\left(\frac{4}{2}\right)\right] 
\label{eq:204} \\
&= f_{\text{Ta}}\cos(0) + f_{\text{Pr}}\cos(6\pi) + f_{\text{N1}}\cos(2\pi) + f_{\text{N2}}\cos(4\pi)
\end{eqnarray}

Using $\cos(0) = \cos(2\pi) = \cos(4\pi) = \cos(6\pi) = 1$:

\begin{eqnarray}
F_{204} = f_{\text{Ta}} + f_{\text{Pr}} + 2f_{\text{N}}
\end{eqnarray}
As with the (222) peak, the (204) peak has the large structure factor magnitude, which is consistent with its high intensity observed in Fig. \ref{fig:sym}(c).

\subsubsection{Nitrogen Vacancies}
So far, we have assumed full occupancy for all nitrogen sites and have not considered the presence of deficient nitrogen. We first note that introducing nitrogen vacancies into the obtained crystal structure with the refined atomic positions cannot account for the experimental XRD results. This can be demonstrated by comparing the calculated structure factors of the (102) and (204) reflections with the corresponding experimental observations. In the following, we discuss the results by comparing the experimental and calculated values of the structure factor ratio, $r = F_{204}/F_{102}$.

\paragraph{Nitrogen defect on N1 site}
Let $x_1$ ($0<x_1\leq1$) denote the occupancy of the N1 site. From eqs. (\ref{eq:102}) and (\ref{eq:204}),
\begin{eqnarray}
F_{102} = f_{\text{Ta}} - f_{\text{Pr}} + (1-x_1)f_{\text{N}}\\
F_{204} = f_{\text{Ta}} + f_{\text{Pr}} + (1+x_1)f_{\text{N}}
\label{eq:102_F}
\end{eqnarray}
Then,
\begin{eqnarray}
r(x_1) = \frac{f_{\text{Ta}} + f_{\text{Pr}} + (1+x_1)f_{\text{N}}}{f_{\text{Ta}} - f_{\text{Pr}} + (1-x_1)f_{\text{N}}}.
\end{eqnarray}
The derivative of $r(x_1)$ with respect to $x_1$ is
\begin{eqnarray}
\frac{\partial r}{\partial x_1} = \frac{2f_{\text{N}}(f_{\text{Ta}}+f_{\text{N}})}{(f_{\text{Ta}} - f_{\text{Pr}} + (1-x_1)f_{\text{N}})^2} > 0.
\end{eqnarray}
Thus, $r(x_1)$ is a monotonically increasing function of $x_1$. The experimental value is $r=2.8$, whereas the calculated value satisfies $r(x_1) > r(0)=6.0$ for $0<x_1\leq1$. Therefore, nitrogen vacancies at the N1 site cannot reproduce the experimental value.

\paragraph{Nitrogen defect on N2 site}
Let $x_2$ ($0<x_2\leq1$) denote the occupancy of the N2 site. Following the same procedure as for the N1 site,
\begin{eqnarray}
F_{102} = f_{\text{Ta}} - f_{\text{Pr}} - (1-x_2)f_{\text{N}}\\
F_{204} = f_{\text{Ta}} + f_{\text{Pr}} + (1+x_2)f_{\text{N}}
\end{eqnarray}
Then,
\begin{eqnarray}
r(x_2) = \frac{f_{\text{Ta}} + f_{\text{Pr}} + (1+x_2)f_{\text{N}}}{f_{\text{Ta}} - f_{\text{Pr}} - (1-x_2)f_{\text{N}}}.
\end{eqnarray}
The derivative of $r(x_2)$ with respect to $x_2$ is
\begin{eqnarray}
\frac{\partial r}{\partial x_2} = \frac{-2f_{\text{N}}(f_{\text{Pr}}+f_{\text{N}})}{(f_{\text{Ta}} - f_{\text{Pr}} - (1-x_2)f_{\text{N}})^2} < 0.
\end{eqnarray}
Thus, $r(x_2)$ is a monotonically decreasing function of $x_2$. The experimental value is $r=2.8$, whereas the calculated value satisfies $r(x_2)\geq r(1)=7.7$ for $0<x_2\leq1$. Therefore, nitrogen vacancies at the N2 site also cannot reproduce the experimental value.

Overall, introducing nitrogen vacancies cannot account for the experimentally observed value of $r$.

One possible way to reconcile this discrepancy is to introduce excess nitrogen at interstitial sites. However, there are too many possible interstitial sites to establish a unique structural model. Determining the positions of the excess nitrogen atoms will require further structural studies, such as single-crystal neutron diffraction measurements on \ch{PrTaN2}.

\begin{sidewaystable}
\centering
\begin{tabular}{c c c c c}
\hline
Wyckoff Letter & Multiplicity & Site Symmetry & Coordinates & Impact on $F_{hkl}$ \\
\hline
1a to 1d & 1 & $222$ & $(0,0,0)$, $(0,0,\frac{1}{2})$, etc. & $B = 0$ (acts locally centrosymmetric)\\
2e to 2i & 2 & $2$ & $(x,0,0)$, $(0,y,0)$, $(0,0,z)$, etc. & Reduces number of terms in the sum\\
\hline
\end{tabular}
\caption{Wyckoff positiones of $P222$ and its impact on structure factor $F_{hkl}$}
\label{tbl:wyc_N}
\end{sidewaystable}

\section{Conclusion}
We grew a novel ternary nitride \ch{PrTaN2} thin film on a YAO substrate using a custom-designed molecular beam epitaxy system. Detailed STEM and XRD measurements reveal that the crystal system is orthorhombic. The estimated lattice constants are consistent with the XRD experimental data. Reciprocal space map measurements indicate that the film is essentially strain-free. In selecting substrates, not only lattice mismatch but also surface stability under nitridation conditions should be considered.

For a novel material discovered in thin-film form, determining the chemical formula and crystal structure remains a significant challenge. We systematically ruled out alternative candidate phases in the Pr-Ta-N system. To determine the crystal structure, we developed a fitting procedure based on structure factors. By utilizing constraints from Wyckoff positions, the number of free parameters in fitting Eq.(\ref{eq:F}) to the experimental data can be significantly reduced. Using this approach, we determined the space group $P222$ and atomic coordinates of \ch{PrTaN2}. This study provides a pathway for the exploration of new ternary nitride thin films.

\newpage
\bibliography{PrTaN}

@article{Nelson1945,
   author = {J B Nelson and D P Riley},
   doi = {10.1088/0959-5309/57/3/302},
   issn = {},
   issue = {3},
   journal = {Proc. Phys. Soc.},
   month = {5},
   pages = {160-177},
   title = {An experimental investigation of extrapolation methods in the derivation of accurate unit-cell dimensions of crystals},
   volume = {57},
   url = {},
   year = {1945},
}

@article{Yamamoto2013,
   author = {Hideki Yamamoto and Yoshiharu Krockenberger and Michio Naito},
   doi = {10.1016/j.jcrysgro.2012.12.057},
   issn = {},
   journal = {	J. Cryst. Growth},
   pages = {184-188},
   publisher = {Elsevier},
   title = {Multi-source {MBE} with high-precision rate control system as a synthesis method sui generis for multi-cation metal oxides},
   volume = {378},
   url = {},
   year = {2013},
}

@article{Niewa1998,
   author = {R. Niewa and F. J. DiSalvo},
   doi = {10.1021/cm980137c},
   issn = {},
   issue = {10},
   journal = {Chem. Mater.},
   month = {10},
   pages = {2733-2752},
   publisher = {American Chemical Society},
   title = {Recent Developments in Nitride Chemistry},
   volume = {10},
   url = {},
   year = {1998},
}

@article{DiSalvo1996,
   author = {Francis J DiSalvo and Simon J Clarke},
   doi = {10.1016/S1359-0286(96)80091-X},
   issn = {},
   issue = {2},
   journal = {Curr. Opin. Solid State Mater. Sci.},
   month = {4},
   pages = {241-249},
   title = {Ternary nitrides: a rapidly growing class of new materials},
   volume = {1},
   url = {},
   year = {1996},
}

@article{Sun2019,
   author = {Wenhao Sun and others},
   doi = {10.1038/s41563-019-0396-2},
   issn = {},
   issue = {7},
   journal = {Nat. Mater.},
   month = {7},
   pages = {732-739},
   pmid = {31209391},
   publisher = {Nature Publishing Group},
   title = {{A} map of the inorganic ternary metal nitrides},
   volume = {18},
   year = {2019},
}

@article{Klo2021,
   author = {Simon D. Kloß and Martin L. Weidemann and J. Paul Attfield},
   doi = {10.1002/anie.202108759},
   issn = {15213773},
   issue = {41},
   journal = {Angew. Chem. Int. Ed.},
   month = {10},
   pages = {22260-22264},
   pmid = {34355842},
   publisher = {John Wiley and Sons Inc},
   title = {Preparation of Bulk-Phase Nitride Perovskite LaReN3 and Topotactic Reduction to LaNiO2-Type LaReN2},
   volume = {60},
   year = {2021}
}

@Misc{NIST_ASD,
author = {A.~Kramida and {Yu.~Ralchenko} and
J.~Reader and {and NIST ASD Team}},
HOWPUBLISHED = {{NIST Atomic Spectra Database
(ver. 5.12), [Online]. Available:
{\tt{https://physics.nist.gov/asd}} [2026, March 30].
National Institute of Standards and Technology,
Gaithersburg, MD.}},
year = {2024},
}

@article{Sarmiento-Prez2015,
   author = {Rafael Sarmiento-Pérez and Tiago F. T. Cerqueira and Sabine Körbel and Silvana Botti and Miguel A. L. Marques},
   doi = {10.1021/acs.chemmater.5b02026},
   issn = {0897-4756},
   issue = {17},
   journal = {Chem. Mater.},
   month = {9},
   pages = {5957-5963},
   publisher = {American Chemical Society},
   title = {Prediction of Stable Nitride Perovskites},
   volume = {27},
   url = {https://pubs.acs.org/doi/10.1021/acs.chemmater.5b02026},
   year = {2015}
}

@inbook{Pennycook2012,
   author = {Stephen J. Pennycook},
   doi = {10.1002/0471266965.com083.pub2},
   booktitle = {Characterization of Materials},
   month = {10},
   pages = {1-29},
   publisher = {Wiley},
   title = {Scanning Transmission Electron Microscopy},
   year = {2012}
}

@article{Talley2021,
   author = {Kevin R. Talley and Craig L. Perkins and David R. Diercks and Geoff L. Brennecka and Andriy Zakutayev},
   doi = {10.1126/science.abm3466},
   issn = {0036-8075},
   issue = {6574},
   journal = {Science},
   month = {12},
   pages = {1488-1491},
   title = {Synthesis of {LaWN$_3} $nitride perovskite with polar symmetry},
   volume = {374},
   url = {https://www.science.org/doi/10.1126/science.abm3466},
   year = {2021}
}

@article{Cario2001,
   author = {Laurent Cario and Zoltán A. Gál and Thomas P. Braun and Francis J. DiSalvo and Björn Blaschkowski and H.-Jürgen Meyer},
   doi = {10.1006/jssc.2001.9359},
   issn = {00224596},
   issue = {1},
   journal = {J. Sol. Sta. Chem.},
   month = {11},
   pages = {90-95},
   publisher = {Academic Press Inc.},
   title = {Ln3T2N6 (Ln=La, Ce, Pr; T=Ta, Nb), a New Family of Ternary Nitrides Isotypic to a High Tc Cuprate Superconductor},
   volume = {162},
   url = {https://linkinghub.elsevier.com/retrieve/pii/S0022459601993594},
   year = {2001}
}

@article{Choi2025,
   author = {Songhee Choi and Qiao Jin and Xian Zi and Dongke Rong and Jie Fang and Jinfeng Zhang and Qinghua Zhang and Wei Li and Shuai Xu and Shengru Chen and Haitao Hong and Cui Ting and Qianying Wang and Gang Tang and Chen Ge and Can Wang and Zhiguo Chen and Lin Gu and Qian Li and Lingfei Wang and Shanmin Wang and Jiawang Hong and Kuijuan Jin and Er-Jia Guo},
   doi = {10.1126/sciadv.adu6698},
   issn = {2375-2548},
   issue = {32},
   journal = {Sci. Adv.},
   month = {8},
   title = {A single-phase epitaxially grown ferroelectric perovskite nitride},
   volume = {11},
   pages = {eadu6698},
   url = {https://www.science.org/doi/10.1126/sciadv.adu6698},
   year = {2025}
}

@article{Rusevich2021,
   author = {L. L. Rusevich and M. Tyunina and E. A. Kotomin and N. Nepomniashchaia and A. Dejneka},
   doi = {10.1038/s41598-021-02751-9},
   issn = {2045-2322},
   issue = {1},
   journal = {Sci. Rep.},
   month = {12},
   pages = {23341},
   publisher = {Nature Research},
   title = {The electronic properties of SrTiO3-δ with oxygen vacancies or substitutions},
   volume = {11},
   url = {https://www.nature.com/articles/s41598-021-02751-9},
   year = {2021}
}

@article{Choi2011,
   author = {Minseok Choi and Fumiyasu Oba and Isao Tanaka},
   doi = {10.1103/PhysRevB.83.214107},
   issn = {1098-0121},
   issue = {21},
   journal = {Phys. Rev. B},
   month = {6},
   pages = {214107},
   title = {Hybrid density functional study of oxygen vacancies in KTaO      3     and NaTaO      3    },
   volume = {83},
   url = {https://link.aps.org/doi/10.1103/PhysRevB.83.214107},
   year = {2011}
}

@article{Fu2018,
   author = {Mingxue Fu and Tingyu Liu and Xiaoxiao Lu and Jing Li and Zhiming Ma},
   doi = {10.1016/j.commatsci.2017.09.016},
   issn = {09270256},
   journal = {Computational Materials Science},
   month = {1},
   pages = {127-132},
   publisher = {Elsevier B.V.},
   title = {First-principles optical spectra for the oxygen vacancy in YAlO3 crystal},
   volume = {141},
   url = {https://linkinghub.elsevier.com/retrieve/pii/S0927025617304962},
   year = {2018}
}

@article{Weidemann2024,
   author = {M. Weidemann and D. Werhahn and C. Mayer and S. Kläger and C. Ritter and P. Manuel and J. P. Attfield and Simon D. Kloß},
   doi = {10.1038/s41557-024-01558-1},
   issn = {1755-4330},
   journal = {Nat. Chem.},
   month = {6},
   title = {High-pressure synthesis of Ruddlesden-Popper nitrides},
   pages = {1723-1731},
   url = {https://www.nature.com/articles/s41557-024-01558-1},
   year = {2024}
}

@article{Broll1995,
  author  = {Broll, S. and Jeitschko, W.},
  journal = {Z. Naturforsch.},
  volume  = {50},
  pages   = {905--912},
  year    = {1995}
}

@article{Seto2022,
   author = {Y. Seto and M. Ohtsuka},
   doi = {10.1107/S1600576722000139},
   issn = {1600-5767},
   issue = {2},
   journal = {J. Appl. Cryst.},
   month = {4},
   pages = {397-410},
   title = {ReciPro : free and open-source multipurpose crystallographic software integrating a crystal model database and viewer, diffraction and microscopy simulators, and diffraction data analysis tools},
   volume = {55},
   year = {2022}
}

@inbook{Looijenga-Vos2006,
   author = {A. Looijenga-Vos and M. J. Buerger},
   city = {Chester, England},
   doi = {10.1107/97809553602060000506},
   booktitle = {International Tables for Crystallography},
   month = {10},
   pages = {44-54},
   publisher = {International Union of Crystallography},
   title = {Space-group determination and diffraction symbols},
   url = {https://xrpp.iucr.org/cgi-bin/itr?url_ver=Z39.88-2003&rft_dat=what%3Dchapter%26volid%3DAb%26chnumo%3D3o1%26chvers%3Dv0001},
   year = {2006}
}

@inbook{Birkholz2005,
   author = {Mario Birkholz and Christoph Genzel},
   doi = {10.1002/3527607595.ch6},
   booktitle = {Thin Film Analysis by X-Ray Scattering},
   month = {10},
   pages = {239-295},
   publisher = {Wiley},
   title = {Residual Stress Analysis},
   year = {2005}
}

@article{Zakutayev2024,
   author = {Andriy Zakutayev and Matthew Jankousky and Laszlo Wolf and Yi Feng and Christopher L. Rom and Sage R. Bauers and Olaf Borkiewicz and David A. LaVan and Rebecca W. Smaha and Vladan Stevanovic},
   doi = {10.1038/s44160-024-00643-0},
   issn = {27310582},
   journal = {Nat. Synth.},
   month = {12},
   pages = {1471-1480},
   volume = {3},
   publisher = {Nature Publishing Group},
   title = {Synthesis pathways to thin films of stable layered nitrides},
   year = {2024}
}

@article{Grosso2023,
   author = {Bastien F. Grosso and Daniel W. Davies and Bonan Zhu and Aron Walsh and David O. Scanlon},
   doi = {10.1039/D3SC02171H},
   issn = {2041-6520},
   issue = {34},
   journal = {Chemical Science},
   month = {8},
   pages = {9175-9185},
   publisher = {Royal Society of Chemistry},
   title = {Accessible chemical space for metal nitride perovskites},
   volume = {14},
   url = {http://xlink.rsc.org/?DOI=D3SC02171H},
   year = {2023}
}

@article{Flores-Livas2019,
   author = {José A. Flores-Livas and R. Sarmiento-Pérez and Silvana Botti and Stefan Goedecker and Miguel A.L. Marques},
   doi = {10.1088/2515-7639/ab083e},
   issn = {25157639},
   issue = {2},
   journal = {J. Phys. Mater.},
   month = {4},
   publisher = {IOP Publishing Ltd},
   title = {Rare-earth magnetic nitride perovskites},
   volume = {2},
   pages = {025003},
   year = {2019}
}

@article{Cordes2019,
   author = {Niklas Cordes and Robin Niklaus and Wolfgang Schnick},
   doi = {10.1021/acs.cgd.9b00357},
   issn = {1528-7483},
   issue = {6},
   journal = {Cryst. Growth Des.},
   month = {6},
   pages = {3484-3490},
   publisher = {American Chemical Society},
   title = {Ammonothermal Crystal Growth of {ATaN$_2$ with A = Na, K, Rb, and Cs} and Their Optical and Electronic Properties},
   volume = {19},
   url = {https://pubs.acs.org/doi/10.1021/acs.cgd.9b00357},
   year = {2019}
}

@article{Tomaschko2012,
   author = {J. Tomaschko and V. Leca and T. Selistrovski and S. Diebold and J. Jochum and R. Kleiner and D. Koelle},
   doi = {10.1103/PhysRevB.85.024519},
   issn = {1098-0121},
   issue = {2},
   journal = {Phys. Rev. B},
   month = {1},
   pages = {024519},
   title = {Properties of the electron-doped infinite-layer superconductor {Sr_{1-x}La_xCuO_2} epitaxially grown by pulsed laser deposition},
   volume = {85},
   url = {https://link.aps.org/doi/10.1103/PhysRevB.85.024519},
   year = {2012}
}

@article{Mercey1995,
   author = {B. Mercey and A Gupta and M. Hervieu and B. Raveau},
   doi = {10.1006/jssc.1995.1218},
   issn = {00224596},
   issue = {2},
   journal = {J. Solid State Chem.},
   month = {5},
   pages = {300-306},
   title = {A New Ordered Oxygen-Deficient Perovskite Sm2Sr6Cu8O17+δ: HREM Study of PLD Thin Films},
   volume = {116},
   url = {https://linkinghub.elsevier.com/retrieve/pii/S0022459685712182},
   year = {1995}
}

@article{Adachi1992,
   author = {H. Adachi and T. Satoh and Y. Ichikawa and K. Setsune and K. Wasa},
   doi = {10.1016/0921-4534(92)90131-U},
   issn = {09214534},
   issue = {1-2},
   journal = {Physica C},
   month = {6},
   pages = {14-16},
   title = {Superconducting (Sr, Nd) CuOy thin films with infinite-layer structure},
   volume = {196},
   url = {https://linkinghub.elsevier.com/retrieve/pii/092145349290131U},
   year = {1992}
}

@Misc{NIST_MAC,
author = {Hubbell, J.H. and Seltzer, S.M. },
HOWPUBLISHED = {Tables of X-Ray Mass Attenuation Coefficients and Mass Energy-Absorption Coefficients (version 1.4). [Online] Available: {http://physics.nist.gov/xaamdi}[2026, March 30].
National Institute of Standards and Technology, Gaithersburg, MD.},
year = {2004},
}

\end{document}